\documentclass[twocolumn,showpacs,preprintnumbers,amsmath,amssymb]{revtex4}
\usepackage{graphicx}
\usepackage{dcolumn}
\usepackage{bm}
\usepackage{rotating}
\usepackage{color}

\begin{document}


\title{1+1+1 flavor QCD+QED simulation at the physical point}

\author{
S.~Aoki$^{1,2}$, K.~-I.~Ishikawa$^3$,
N.~Ishizuka$^{1,2}$,
K.~Kanaya$^2$, Y.~Kuramashi$^{1,2,4}$,
Y.~Nakamura$^4$, Y.~Namekawa$^1$,
M.~Okawa$^3$, Y.~Taniguchi$^{1,2}$,
A.~Ukawa$^{1,2}$, N.~Ukita$^1$ and T.~Yoshi\'e$^{1,2}$\\
(PACS-CS Collaboration)
}

\affiliation{
$^1$~Center for Computational Sciences, University of Tsukuba,
Tsukuba, Ibaraki 305-8577, Japan\\
$^2$~Graduate School of Pure and Applied Sciences, University of
Tsukuba,
Tsukuba, Ibaraki 305-8571, Japan\\
$^3$~Graduate School of Science, Hiroshima University,
Higashi-Hiroshima, Hiroshima 739-8526, Japan\\
$^4$~RIKEN Advanced Institute for Computational Science,
Kobe, Hyogo 650-0047, Japan
}
%
\collaboration{ PACS-CS Collaboration }

\date{\today}

\begin{abstract}
We present the results of 1+1+1 flavor QCD+QED simulation at the  
physical point, in which the dynamical quark effects in QED and the up-down quark mass difference
are incorporated by the reweighting technique. The physical quark masses together with 
the lattice spacing are determined with $m_{\pi^+}$, $m_{K^+}$, $m_{K^0}$ and $m_{\Omega^-}$ as physical inputs. 
Calculations are carried out using a set of 2+1 flavor QCD configurations near the physical point generated by the non-perturbatively $O(a)$-improved Wilson quark action and the Iwasaki gauge action at $\beta=1.9$ on a $32^3\times 64$ lattice. We evaluate the values of the up, down and strange quark masses individually with non-perturbative QCD renormalization.

\end{abstract}

\maketitle

\section{Introduction}
The ultimate goal of lattice QCD is to understand the non-perturbative dynamics of strong interactions at the physical values of the up, down and strange quark masses. Recently we have achieved this goal assuming artificial isospin symmetry with degenerate up and down quark masses $m_{\rm u} = m_{\rm d}$\cite{PACS-CS_1, PACS-CS_2}. In nature, however, the isospin symmetry is broken due to the up-down quark mass difference $m_{\rm u} \neq m_{\rm d}$ and their electric charge difference. Their effects are observed in mass splittings among isospin multiplets of light hadrons, {\it e.g.} $m_{K^0}-m_{K^\pm}$, $m_{n}-m_{p}$. The magnitude of splittings is tiny yet important since, {\it e.g., } it is this difference which guarantees the stability of proton. Thus, our next step should be 1+1+1 QCD+QED lattice simulation at the physical point incorporating the isospin breaking effects.

A pioneering lattice QCD+QED simulation was carried out in Refs.~\cite{Duncan_1, Duncan_2} employing quenched approximation both for QCD and QED, where the photon fields and the gluon fields are independently generated by conventional Monte Carlo methods and they are superimposed. This work was followed by several studies in quenched QED on quenched and dynamical SU(3) gauge configurations\cite{Namekawa}-\cite{Glaessle}. These studies have  shown that the mass splittings among the isospin multiplets of hadrons came out qualitatively consistent with the experimental ones.  Nevertheless, dynamical quark effects in QED should be properly evaluated to obtain convincing results. An inclusion of dynamical quark effects in QED was first attempted in Ref.~\cite{Duncan_3} by calculating the ratio of the quark determinant of QCD+QED to that of QCD on rather small lattices using the reweighting technique.  
Recently results for the low-energy constants including dynamical quark effects both in QCD and QED are reported\cite{Ishikawa}. 

In this article, we present results for 1+1+1 flavor QCD+QED simulation at the physical point. The photon fields are superimposed on the 2+1 flavor QCD configurations near the physical point generated by the PACS-CS Collaboration\cite{PACS-CS_2} employing the non-perturbatively $O(a)$-improved Wilson quark action and the Iwasaki gauge action at $\beta=1.90$, corresponding to the lattice spacing of $a\sim0.1$ fm, on a $32^3\times 64$ lattice. We employ the reweighting technique to incorporate the dynamical quark effects in QED\cite{Duncan_3} and adjust the up and down quark masses to their physical values independently\cite{PACS-CS_2, Hasenfratz, RBC_UKQCD}. We make several improvements on the generation of photon fields and evaluation of the reweighting factors. 

This paper is organized as follows. Section~\ref{sec:reweighting} is devoted to explain the generation of the photon fields and the calculational method of the reweighting factors to incorporate the dynamical quark effects in QED. In Sec.~\ref{sec:result} we present the results for 1+1+1 flavor QCD+QED simulation at the physical point. Our conclusions are summarized in Sec.~\ref{sec:conclusion}.    

\section{Reweighting factors for full QED}
\label{sec:reweighting}
Given a quenched U(1) gauge configuration superimposed on a 2+1 flavor QCD configuration, 
the reweighting factor is evaluated with a set of independent Gaussian random noises $\eta_i$ $(i=1,\dots,N_\eta)$ :
\begin{eqnarray}
 {\rm det}[W_{\rm uds}] &=&
 \left[\lim_{N_\eta\rightarrow \infty}
 \frac{1}{N_\eta} \sum_{i=1}^{N_\eta} 
  {\rm e}^{-\vert W_{\rm uds}^{-1}\eta_i\vert^2 +\vert \eta_i\vert^2}\right]^{\frac{1}{2}},
 \label{eq:rwfactor}\\
 W_{\rm uds} &=& \prod_{q={\rm u,d,s}}\frac{D(e_{\rm ph}Q_q, \kappa_q^*)}{D(0, \kappa_q)}.
\label{eq:w_uds}
\end{eqnarray}
%
where $D(e_{\rm ph}Q_q, \kappa_q^*)$ is the Wilson-Dirac matrix with a hopping parameter $\kappa_q^*$ corresponding to the physical point and the physical electric charge $e_{\rm ph}=\sqrt{{4\pi}/{137}}$ multiplied by a fraction $Q_q$ specific to each quark flavor. 
We employ the determinant breakup technique in terms of both the electric charge and the hopping parameters to reduce the noise fluctuations \cite{PACS-CS_2, Hasenfratz, RBC_UKQCD}.
It is clear that the stochastic evaluation of det$[W_q]$ works efficiently when the ratio $W_q=D(e_{\rm ph}Q_q, \kappa_q^*)/D(0, \kappa_q)$ is close to the unit matrix, for which  local fluctuations of the photon field should be suppressed.   
We employ the following procedure to prepare the photon field.

We first generate the photon field $b_{\mu}(n)$ on a $64^3\times 128$ lattice, which is twice finer than the QCD lattice of $32^3\times 64$, employing a non-compact pure gauge action in the Coulomb gauge and an appropriate treatment of zero modes\cite{Kogut,Duncan_1,Blum_1,Hayakawa}:
\begin{eqnarray}
 S_{\rm photon} &=&
 \sum_{n}\sum_{\mu,\nu}
  \frac{1}{4}(\partial_{\mu}b_{\nu}(n)-\partial_{\nu}b_{\mu}(n))^2 \nonumber \\
  &&+\sum_{n}\sum_{\mu,\nu} c_1(\partial_{\mu}(\partial_{\mu}b_{\nu}(n)-\partial_{\nu}b_{\mu}(n)))^2
\end{eqnarray}
with 
\begin{eqnarray}
 \partial_{\mu}b_{\nu}(n) &=& b_{\nu}(n+\hat\mu/2) - b_{\nu}(n),
\end{eqnarray}
%
where $\hat{\mu}$ is the unit vector in $\mu$ direction on the QCD lattice.
The coefficient $c_1=-0.646$ is chosen such that 
the averaged action density on the QCD lattice defined by 
\begin{eqnarray}
\left\langle \frac{1}{4}(\nabla_{\mu}B_{\nu}(N)-\nabla_{\nu}B_{\mu}(N))^2\right\rangle
\end{eqnarray}
with 
\begin{eqnarray}
B_{\mu}(N=n/2) &=& b_{\mu}(n)+b_{\mu}(n+\hat\mu/2),\\ 
\nabla_{\mu}B_{\nu}(N) &=& B_{\nu}(N+\hat\mu) - B_{\nu}(N)
\end{eqnarray}
is numerically equal to that of the photon field generated with $c_1=0$ on the QCD lattice. 
To reduce local fluctuations we average the photon fields over independent paths inside the $2\times 2\times 2\times 2$ hypercube on the QED lattice embedded in the unit hypercube on the QCD lattice:
\begin{widetext}
\begin{eqnarray}
\bar{B}_{\mu}(N=n/2) &=& 
\frac{1}{27}[ (b_{\mu}(n) + b_{\mu}(n+\hat{\mu}/2)) \nonumber\\ \nonumber
&&+\sum_{\nu\neq\mu}\sum_{\nu_s=\pm\nu}
(
b_{\nu_s}(n)
+ b_{\mu}(n+\hat{\nu_s}/2) 
+ b_{\mu}(n+\hat{\nu_s}/2+\hat{\mu}/2)
- b_{\nu_s}(n+\hat{\mu}) ) \\ \nonumber 
&&+\frac{1}{2}\sum_{\rho\neq\nu\neq\mu}\sum_{\rho_s=\pm\rho}\sum_{\nu_s=\pm\nu}
(
b_{\rho_s}(n)
+b_{\nu_s}(n+\hat{\rho}_s/2) \\ \nonumber &&\ \ \ \ \ \ \ \ 
+b_{\mu}(n+\hat{\rho}_s/2+\hat{\nu}_s/2)
+ b_{\mu}(n+\hat{\rho}_s/2+\hat{\nu}_s/2+\hat{\mu}/2) \\ \nonumber &&\ \ \ \ \ \ \ \ 
-b_{\rho_s}(n+\hat{\mu})
-b_{\nu_s}(n+\hat{\rho}_s/2+\hat{\mu})
) \\ \nonumber 
&&+\frac{1}{6}\sum_{\sigma\neq\rho\neq\nu\neq\mu}
\sum_{\sigma_s=\pm\sigma}\sum_{\rho_s=\pm\rho}\sum_{\nu_s=\pm\nu}
(
b_{\sigma_s}(n)
+b_{\rho_s}(n+\hat{\sigma}_s/2)
+b_{\nu_s}(n+\hat{\sigma}_s/2+\hat{\rho}_s/2)
 \\ \nonumber &&\ \ \ \ \ \ \ \ 
+b_{\mu}(n+\hat{\sigma}_s/2+\hat{\rho}_s/2+\hat{\nu}_s/2)
+b_{\mu}(n+\hat{\sigma}_s/2+\hat{\rho}_s/2+\hat{\nu}_s/2+\hat{\mu}/2)
 \\  &&\ \ \ \ \ \ \ \ 
-b_{\sigma_s}(n+\hat{\mu})
-b_{\rho_s}(n+\hat{\sigma}_s/2+\hat{\mu})
-b_{\nu_s}(n+\hat{\sigma}_s/2+\hat{\rho}_s/2+\hat{\mu})
)
].
\end{eqnarray}
\end{widetext}
Finally, we construct the compact link variable $U_{\mu}^q = {\rm exp}({ie_{\rm ph}Q_q\bar{B}_{\mu}})$ with $e_{\rm ph}Q_q$ the physical electric charge of each quark. 

We have checked that the averaging procedure does not cause variation of the electric charge by calculating the electromagnetic mass splittings among charged and neutral pseudoscalar mesons in the flavor non-singlet sector: Results obtained  with the averaged photon field are consistent with those with the conventional ones within error bars. We have also found that this procedure sizably reduces the additive quark mass shift due to the QED effects  for the Wilson-clover quark action we employ. 

\begin{figure}[t]
\begin{center}
\begin{tabular}{cc}
\includegraphics[width=80mm,angle=0]{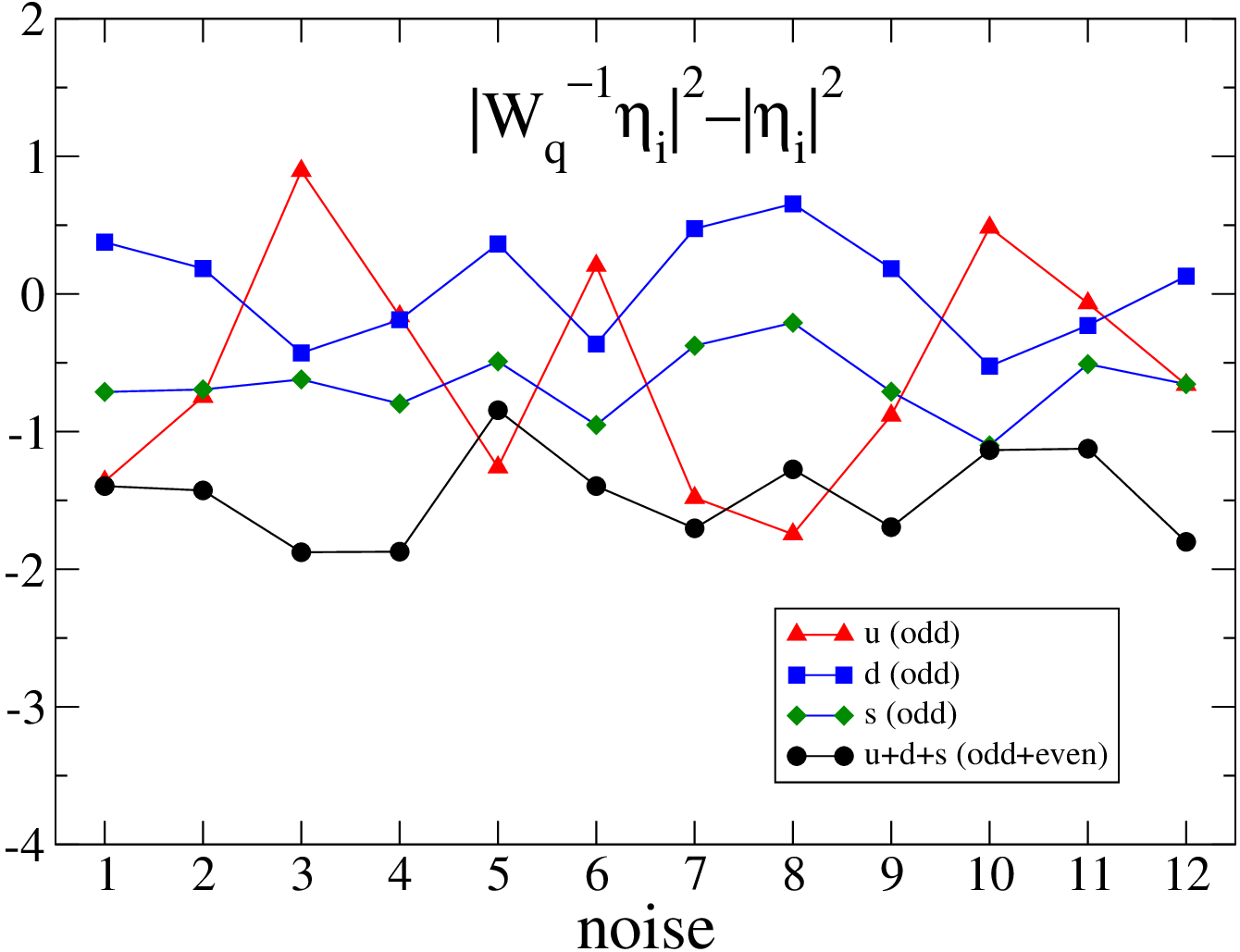}
\end{tabular}
\end{center}
\caption{Noise dependence of the odd part in $e$ for $\vert W_{q}^{-1}\eta_i\vert^2 -\vert \eta_i\vert^2$ ($q={\rm u,d,s}$) with the first breakup interval from $e=0$ to $e_{\rm ph}/400$. Total contributions including both the odd and even parts in $e$ are also presented (black circles). }
\label{fig1}
\end{figure}

Further reduction of the fluctuations in the stochastic evaluation of the reweighting factors is achieved by calculating the combined reweighting factor for all three light quarks with a single set of noise as expressed in Eqs.~(\ref{eq:rwfactor}) and (\ref{eq:w_uds})\cite{Ishikawa}.
It should be noted that the odd terms in $e$ for $\vert W_{\rm uds}^{-1}\eta_i\vert^2 -\vert \eta_i\vert^2$ are not forbidden by the symmetries in the QCD+QED mixed system. The leading contribution in the fluctuations is expected to be of order of $e$. 
Since the total charge for the up, down and strange quarks vanishes, $Q_{\rm u}+Q_{\rm d}+Q_{\rm s}=0$, we can expect a  cancellation of the leading contribution, $O(e)=0$, if the quark masses are identical.  Even though the degeneracy is broken, the partial cancellation of the $O(e)$ terms is expected to reduce the fluctuations in the stochastic evaluation.
In order to illustrate the cancellation we plot the noise dependence of the odd part in $e$ for $\vert W_{q}^{-1}\eta_i\vert^2 -\vert \eta_i\vert^2$ ($q={\rm u,d,s}$) with the first breakup interval $e=0$ to $e_{\rm ph}/400$ in Fig.~\ref{fig1}, where the stochastic evaluation is applied to individual quarks. The odd part in $e$ is extracted from the calculations with $\pm e$.
We observe that the fluctuation for the up quark is opposite to those for the down and strange quarks as a function of the noise, where we expect that the next-to-leading contribution of $O(e^3)$ should be negligibly small.

\section{1+1+1 flavor QCD+QED simulation at the physical point}
\label{sec:result}
\subsection{Parameters}
Our QCD+QED calculation is based on a set of 2+1 flavor QCD configurations near the physical point generated with the non-perturbatively $O(a)$-improved Wilson quark action and the Iwasaki gauge action at $\beta=1.90$ on a $32^3\times 64$ lattice by the PACS-CS Collaboration \cite{PACS-CS_2}. The hopping parameters are $(\kappa_{\rm ud}, \kappa_{\rm s})=(0.13778500, 0.13660000)$. The number of configurations is 80 over 2000 MD time units.  Using the U(1) link variables on a $32^3\times64$ lattice, which are constructed as described in Sec.~\ref{sec:reweighting}, 2+1 flavor QCD partition function is reweighted into 1+1+1 flavor QCD+QED partition function with the hopping parameters ($\kappa_{\rm u}^*,\kappa_{\rm d}^*,\kappa_{\rm s}^*$) and the physical electric charge $e_{\rm ph}$. 
The values of the hopping parameters $(\kappa_{\rm u}^*, \kappa_{\rm d}^*, \kappa_{\rm s}^*)=(0.13787014, 0.13779700, 0.13669510)$ are  determined by $\pi^+, K^+$, $K^0$ and $\Omega^-$ masses as the physical inputs.
We employ 400 determinant breakups for the QED part and 26 for the fine tuning of the hopping parameters to the physical point. Each piece of the divided determinant is evaluated stochastically with 12 sets of noises.
The quark matrix inversions are obtained by a block solver algorithm developed in Ref.~\cite{Nakamura}, which reduce the computational cost by a factor of 3 to 4 compared with non-block solvers.
%

\subsection{Results}

\begin{figure}[t!]
\begin{center}
\begin{tabular}{c}
\includegraphics[width=80mm,angle=0]{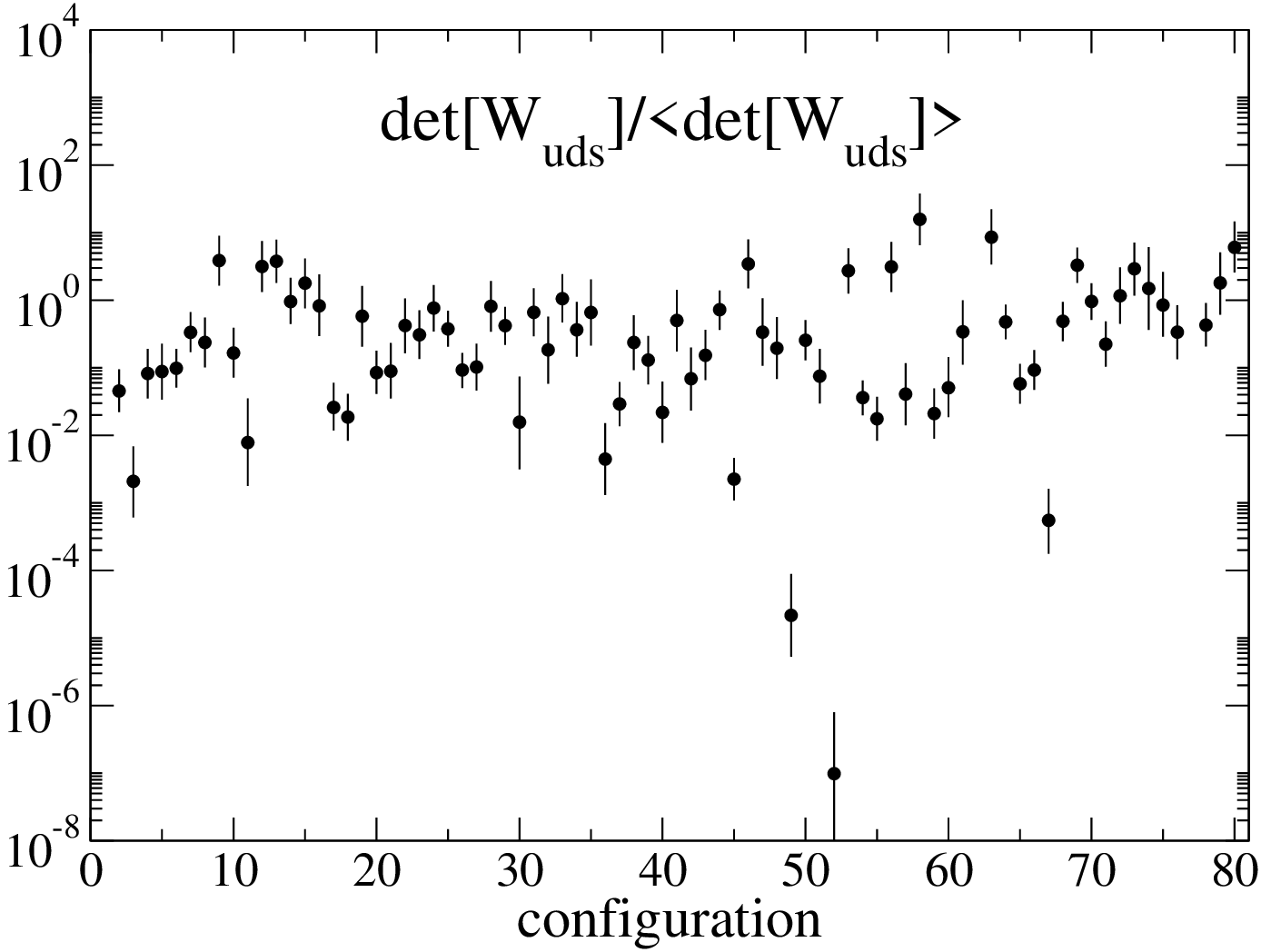}
\end{tabular}
\end{center}
\caption{Configuration dependence of the reweighting factor from 2+1 flavor QCD to 1+1+1 flavor QCD+QED at the physical point.}
\label{fig2}
\end{figure}

In Fig.~\ref{fig2} we plot the configuration dependence  of the reweighting factor from 2+1 flavor QCD to 1+1+1 flavor QCD+QED at the physical point, which are normalized by the configuration average. The largest value is about 10 which is the same order as found in Ref.~\cite{Ishikawa}. 

\begin{table}[t!]
\begin{center}
\renewcommand{\arraystretch}{1.5}
\begin{tabular}{lcc}
\hline\hline
& Our results [MeV] & Experiment [MeV]\\
\hline
$m_{\pi^+}$ & 137.7(8.0) & 139.57018(35) \\
$m_{K^+}$ & 492.3(4.7) & 493.677(16) \\ 
$m_{K^0}$ & 497.4(3.7) & 497.614(24) \\ 
$m_{\Omega^-}$ & input & 1672.45(29) \\ 
\hline\hline
\end{tabular}
\end{center}
\caption{Hadron masses in physical units.}
\label{tab1}
\end{table}%

\begin{table*}[t!]
\setlength{\tabcolsep}{10pt}
\renewcommand{\arraystretch}{1.2}
\centering
\caption{Quark masses and their ratios in comparison with our previous work.
All results are given with the non-perturbative renormalization factors\protect{\cite{PACS-CS_3}}.
The first error is statistical and the second  comes from the renormalization factor, 
while the two errors were combined in the previous work.}
\label{tab:qmass}
\begin{ruledtabular}
\begin{tabular}{lllll}
& This work & Refs.~\protect{\cite{PACS-CS_3,PACS-CS_2}} & Refs.~\protect{\cite{PACS-CS_3,PACS-CS_1}}  \\
\hline
$m^{\overline{\rm MS}}_{\rm u}$ [MeV]  & 2.57(26)(07) & $\cdots$ & $\cdots$  \\
$m^{\overline{\rm MS}}_{\rm d}$  [MeV]  & 3.68(29)(10) & $\cdots$ & $\cdots$  \\
$m^{\overline{\rm MS}}_{\rm s}$  [MeV]  & 83.60(58)(2.23) & 86.7(2.3) & 87.7(3.1) \\
$m^{\overline{\rm MS}}_{\rm ud}$  [MeV]  & 3.12(24)(08) & 2.78(27)  &  3.05(12) \\
$m_{\rm u}/m_{\rm d}$ &  0.698(51)  & $\cdots$ &  $\cdots$  \\
$m_{\rm s}/m_{\rm ud}$ &  26.8(2.0)  & 31.2(2.7) & 28.78(40)  \\
\end{tabular}
\end{ruledtabular}
\end{table*}

\begin{figure}[t!]
\begin{center}
\begin{tabular}{c}
\includegraphics[width=70mm,angle=0]{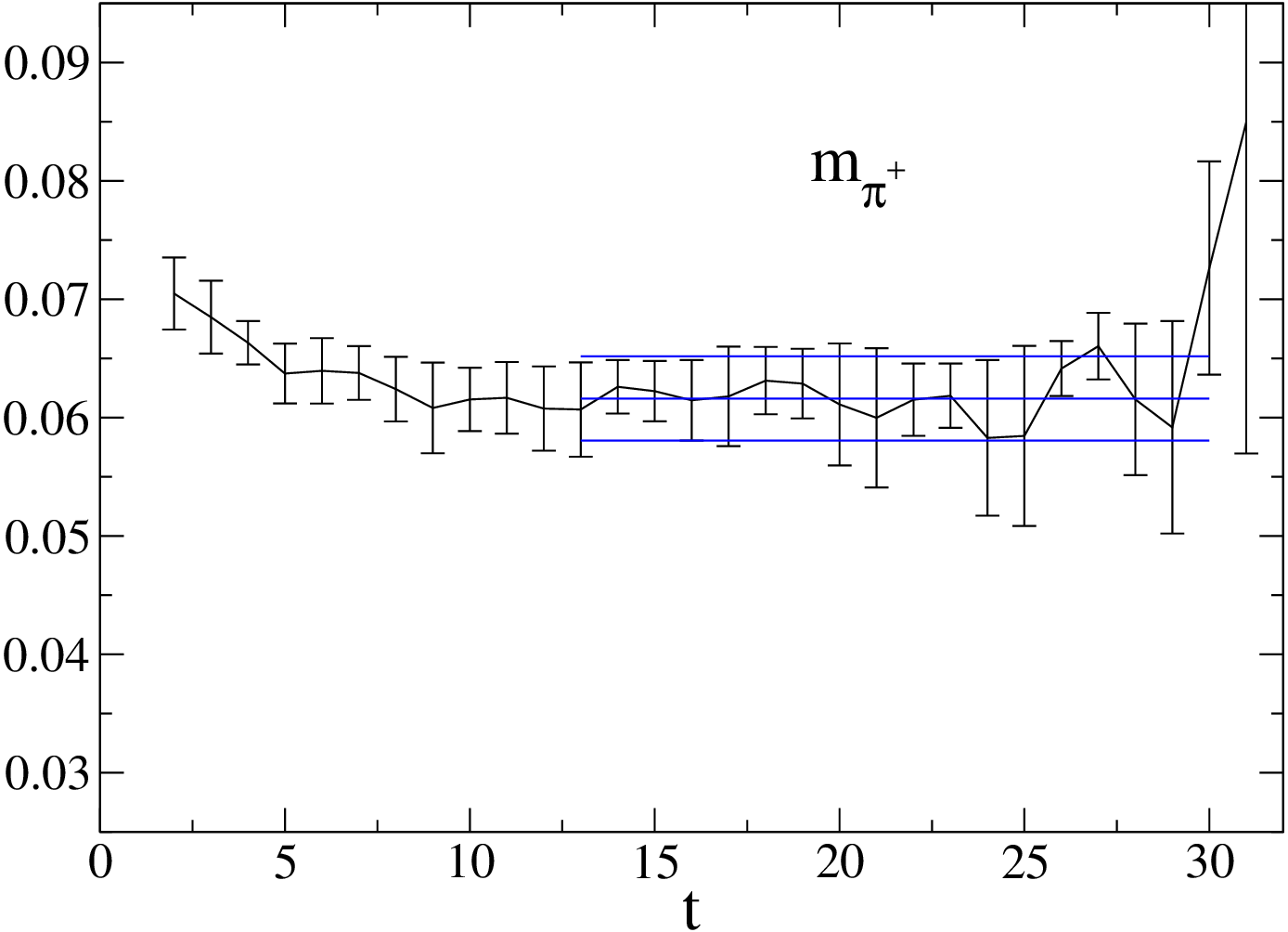}
\\
\includegraphics[width=70mm,angle=0]{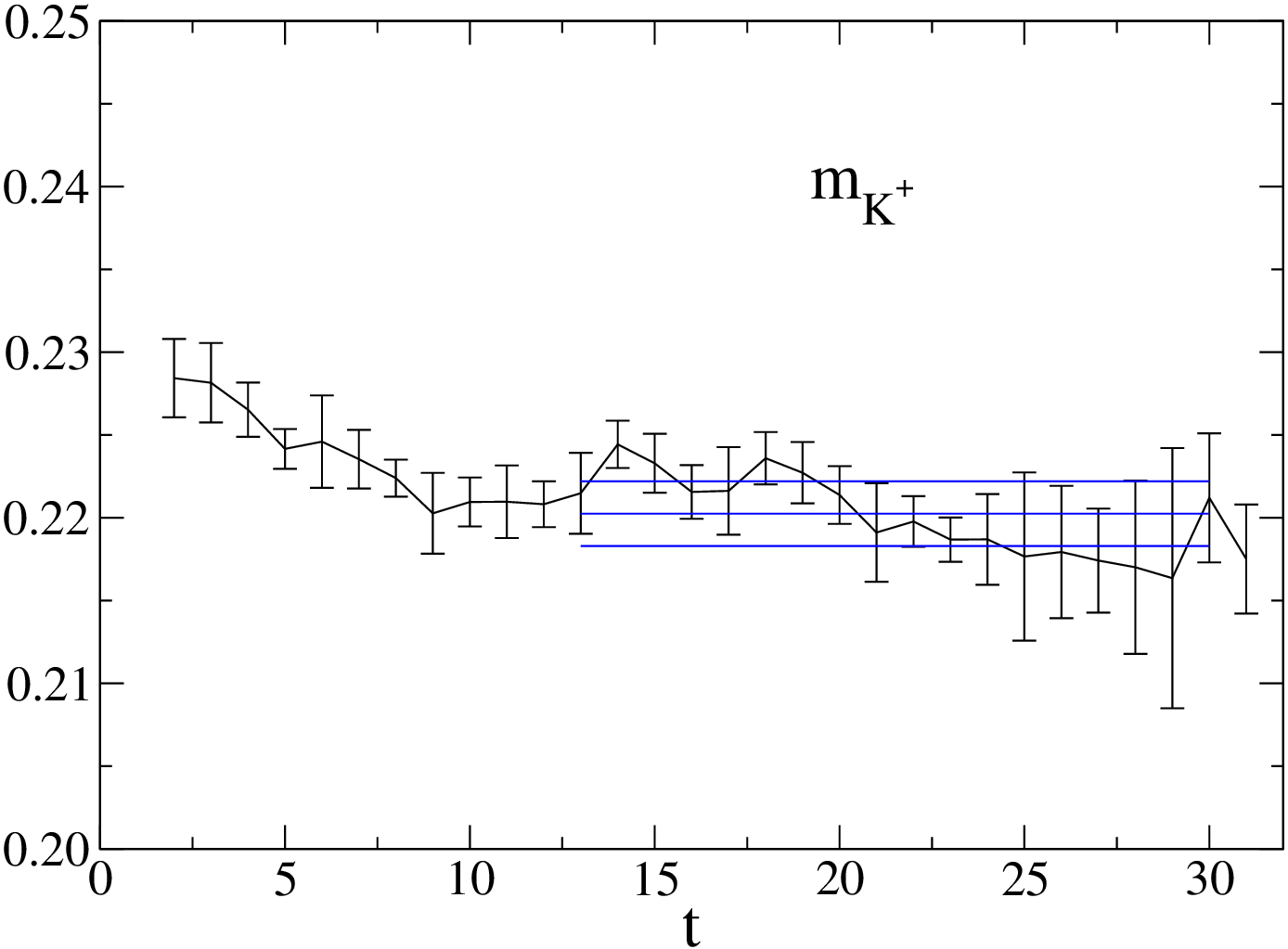}
\\
\includegraphics[width=70mm,angle=0]{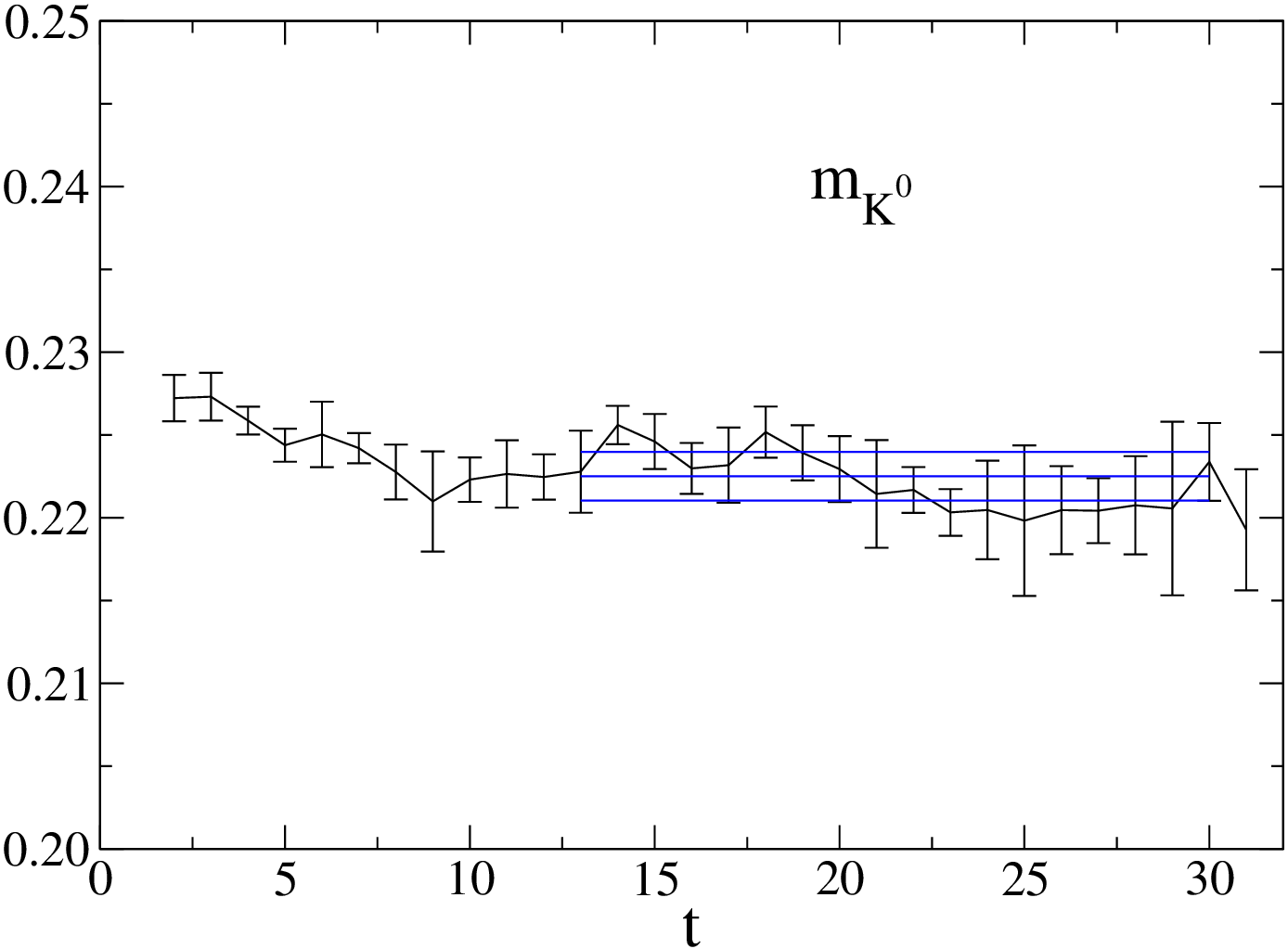}
\\
\includegraphics[width=70mm,angle=0]{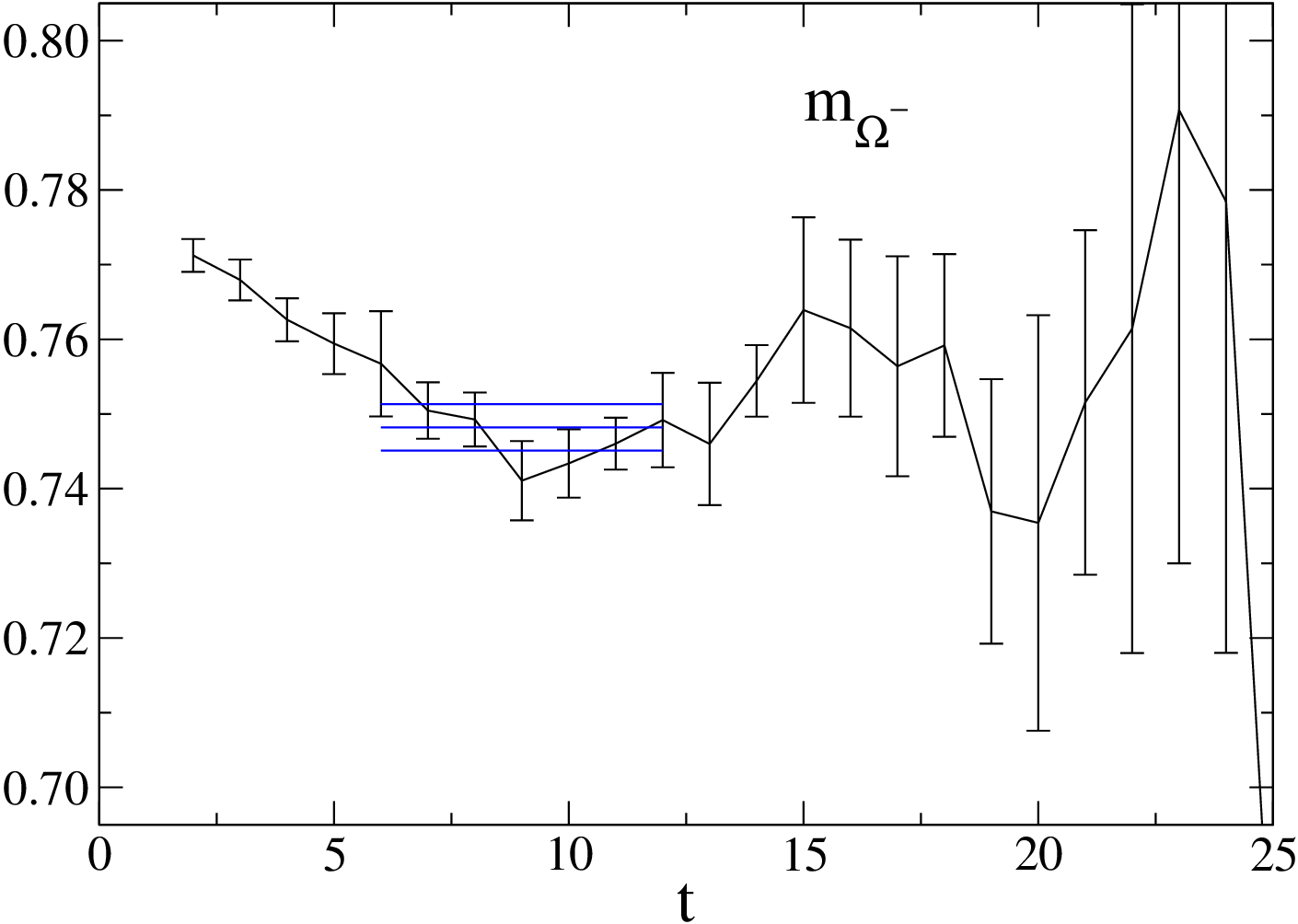}
\end{tabular}
\end{center}
\caption{Effective mass plots of $m_{\pi^+}, m_{K^+}$, $m_{K^0}$ and $m_{\Omega^-}$. Blue horizontal bars represent the fit results with one standard deviation error band.}
\label{fig3}
\end{figure}

\begin{figure}[t!]
\begin{center}
\begin{tabular}{c}
\includegraphics[width=70mm,angle=0]{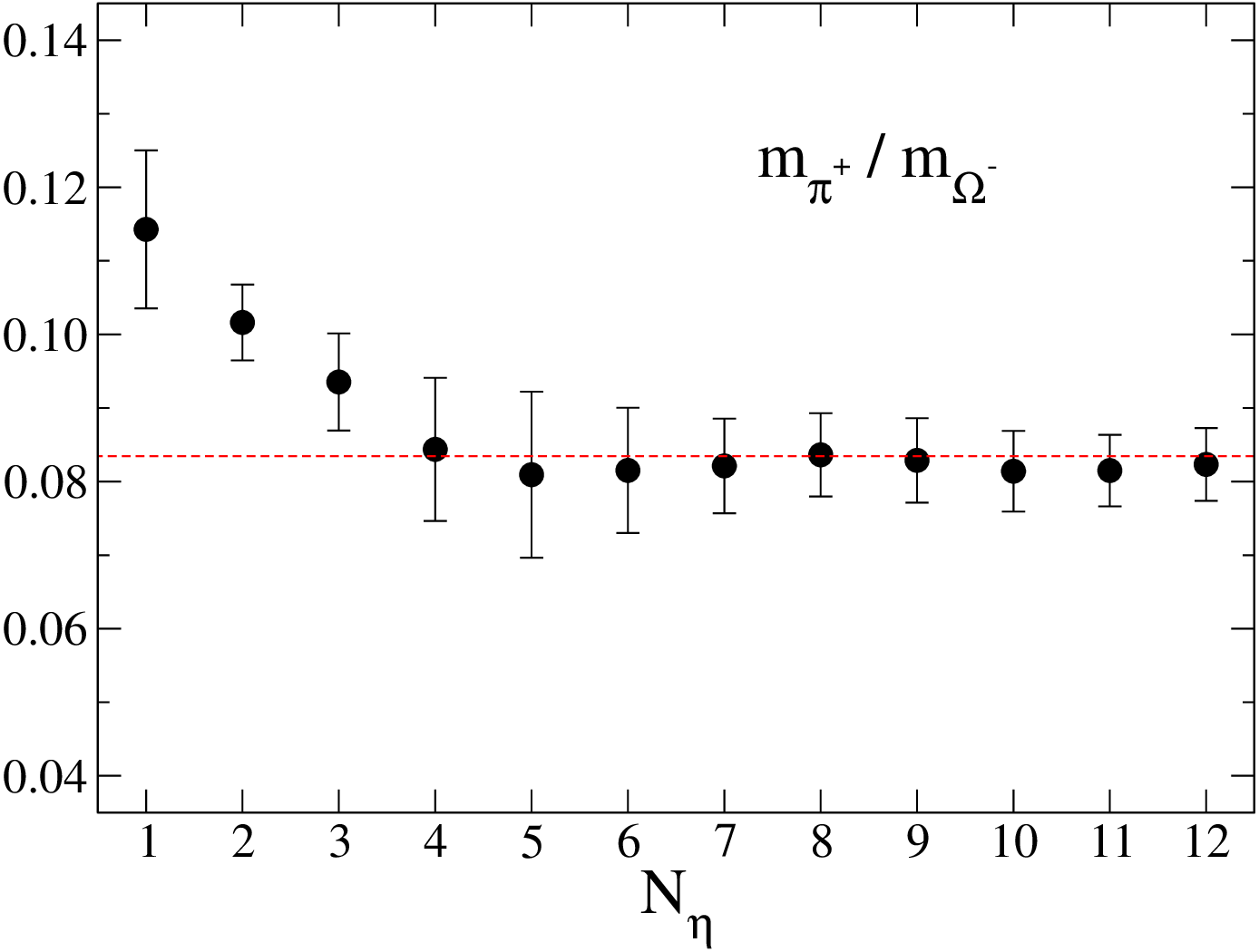}
\\
\includegraphics[width=70mm,angle=0]{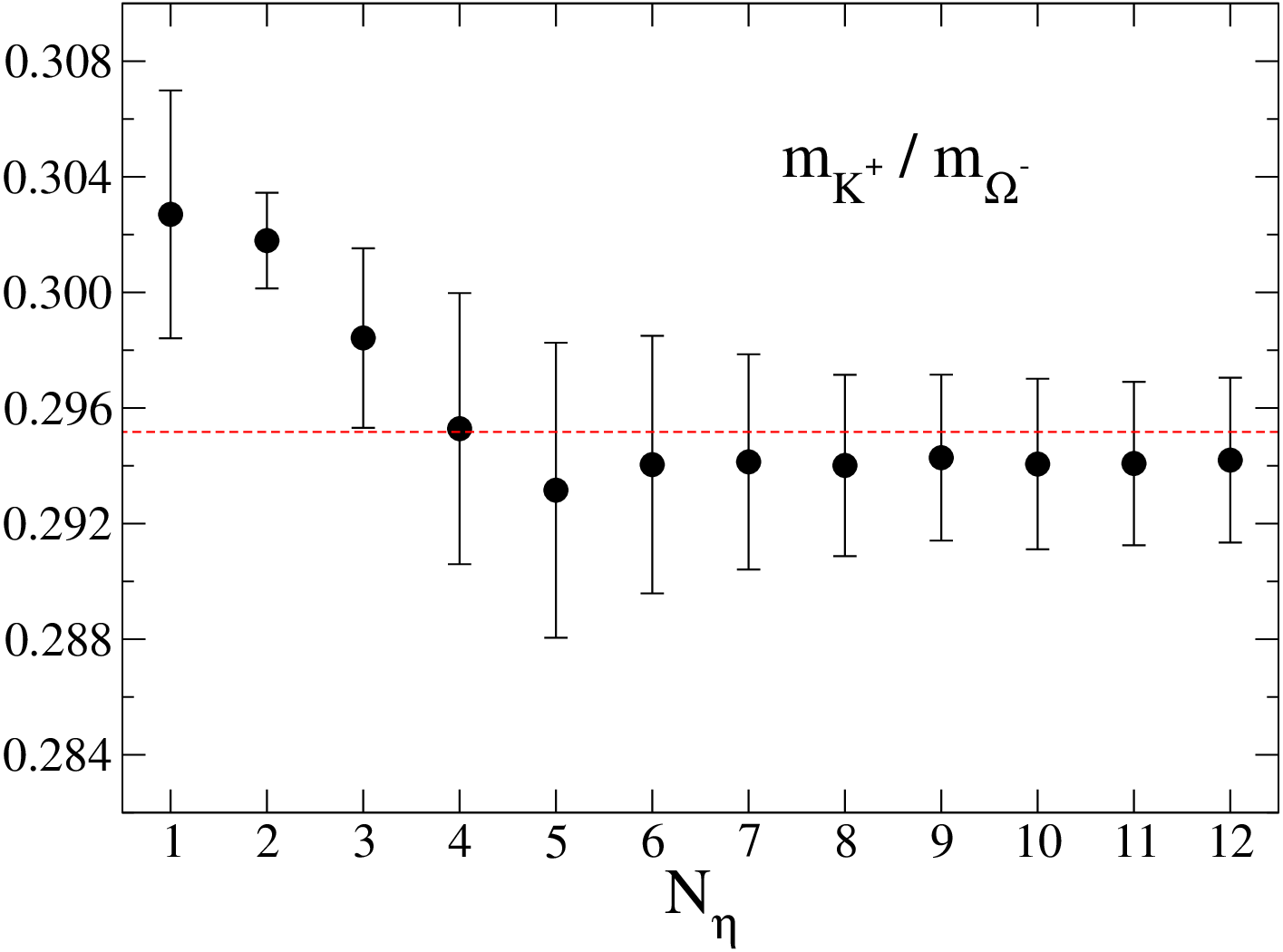}
\\
\includegraphics[width=70mm,angle=0]{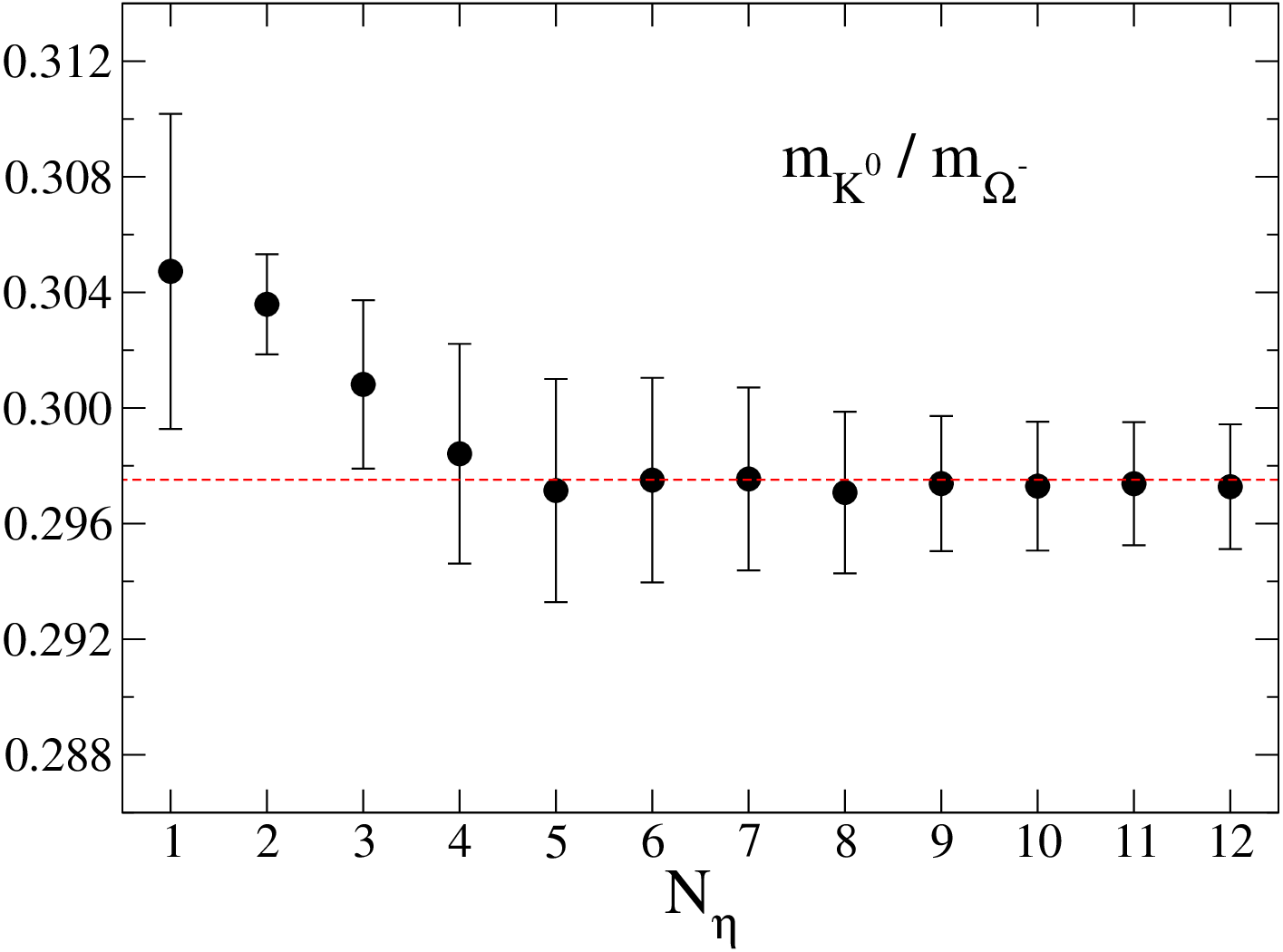}
\end{tabular}
\end{center}
\caption{Hadron mass ratios ${m_{\pi^+}}/{m_{\Omega^-}}$, ${m_{K^+}}/{m_{\Omega^-}}$ and ${m_{K^0}}/{m_{\Omega^-}}$ (black filled circles) as a function of the number of noise for each determinant breakup. Red broken lines indicate the experimental values.}
\label{fig4}
\end{figure}

\begin{figure}[t]
\begin{center}
\begin{tabular}{c}
\includegraphics[width=75mm,angle=0]{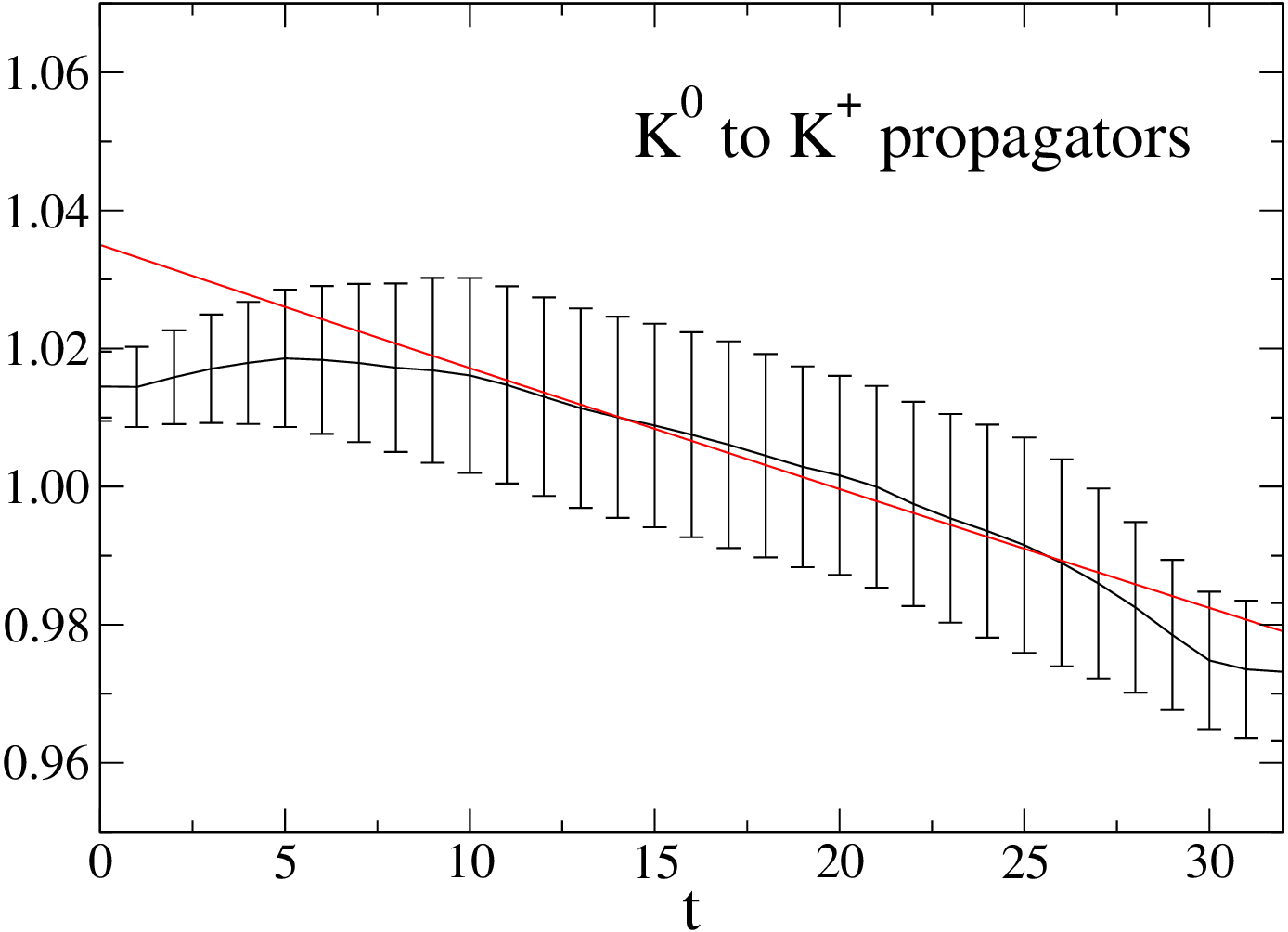}
\end{tabular}
\end{center}
\caption{Ratio of $K^0$ to $K^+$ propagators to detect the mass difference $m_{K^0}-m_{K^+}$. Our results (black symbol) are consistent with the expected slope from the experimental value of $m_{K^0}-m_{K^+}$ (red line).}
\label{fig5}
\end{figure}

We calculate the hadron correlators with point and smeared sources and a local sink employing the operators given in Refs.~\cite{PACS-CS_1, PACS-CS_2}.
The smeared source is constructed with an exponential smearing function 
$\Psi(|{\vec x}|)=A_q\exp(-B_q|{\vec x}|)$ $(q={\rm u,d,s})$ with $A_{\rm u,d}=1.2, B_{\rm u,d}=0.07$ and 
$A_{\rm s}=1.2, B_{\rm s}=0.18$. To reduce the statistical errors we employ 16 source points located at every 4 time slices with different spatial sites.
Figure~\ref{fig3} shows effective mass plots for $\pi^+, K^+, K^0$ and $\Omega^-$ propagators with the smeared source. The three horizontal lines are extended over the fit ranges, which are $[t_{\rm min},t_{\rm max}]=[13,30]$ for the pseudoscalar mesons and $[6,12]$ for $\Omega^-$ baryon, and show the resulting hadron masses with one standard deviation error band.  The statistical errors are estimated with the jackknife method with a binsize of 100 MD time units. The fit results are summarized in Table~\ref{tab1}. The scale is determined as $a^{-1}=2.235(11)$ GeV using the $\Omega^-$ baryon mass as input. 

In Fig.~\ref{fig4} we show the hadron mass ratios, ${m_{\pi^+}}/{m_{\Omega^-}}$, ${m_{K^+}}/{m_{\Omega^-}}$ and ${m_{K^0}}/{m_{\Omega^-}}$, as a function of the number of noises employed for the stochastic evaluation of each determinant breakup. The ratios quickly reach a plateau around  the experimental values (red broken lines) with small number of noises.  This shows not only that twelve noises are sufficient for evaluation of each determinant breakup but also that we have achieved a successful tuning to the physical point.

Figure \ref{fig5} shows the ratio of $K^0$ to $K^+$ propagators whose time dependence is expected as 
\begin{eqnarray}
 \frac{\left\langle K^0(t)K^0(0) \right\rangle}{\left\langle K^+(t)K^+(0)\right\rangle}
   &\simeq &
    Z\left(1 - (m_{K^0}-m_{K^+})t\right), 
\end{eqnarray}
where we assume $(m_{K^0}-m_{K^+})t \ll 1$. The negative slope indicates $m_{K^0}>m_{K^+}$. The fit result with $[t_{\rm min},t_{\rm max}]=[13,30]$ gives 4.54(1.09)\, MeV, which is consistent with the experimental value 3.937(28)\, MeV \cite{Exp} within the error bar. 

In Table~\ref{tab:qmass} we summarize the up, down and strange quark masses renormalized at $\mu = 2$\, GeV in the continuum $\overline{\rm MS}$ scheme
together with the values from our previous work\cite{PACS-CS_3,PACS-CS_2,PACS-CS_1}.
Note that the present results employ the non-perturbative renormalization factor obtained with the Schr\"odinger functional scheme\cite{PACS-CS_3}, and for a proper comparison, the previous results for the light quark masses in Refs.~\cite{PACS-CS_1,PACS-CS_2} are updated with the use of the non-perturbative renormalization factor in Table 14 of Ref.~\cite{PACS-CS_3}.
The first errors are statistical and the second ones are associated with computation of the non-perturbative renormalization factors. 
We neglect the QED corrections to the renormalization factor, whose contribution would be at a level of 1\% or less. 
A potentially large finite size effects in QED are discussed in Ref.~\cite{Hayakawa}. 
The authors in Ref.~\cite{Blum_2} investigate the finite volume effects on the up, down and strange quark masses in (1.8 fm)$^3$ and (2.7 fm)$^3$ boxes, the latter of which is nearly equal to the spatial volume in this work, employing chiral perturbation theory including electromagnetic interactions. Based on their analyses we estimate the systematic errors for our results due to the finite size corrections of QED to be  $-$13.50\%, $+$2.48\%, $-$0.07\% for the up, down, strange quark masses, respectively.

\section{Conclusions}
\label{sec:conclusion}
We have presented the results for 1+1+1 flavor QCD+QED simulation at the physical point. The dynamical quark effects in QED and the up-down quark mass difference are incorporated by the reweighting technique. We have made several improvements to reduce the fluctuations in the stochastic evaluation of the reweighting factors. The physical values for the up, down and strange quark masses are determined by using $m_{\pi^+}, m_{K^+}, m_{K^0}$ and $m_{\Omega^-}$ as input, and quoted in the $\overline{\rm MS}$ scheme at $\mu=2$GeV.  A direct investigation of the finite size effects of QED by changing the spatial volume is left as a future work.

\begin{acknowledgments}
Numerical calculations for the present work have been carried out under the ``Interdisciplinary
Computational Science Program" in Center for Computational Sciences, University of Tsukuba. 
This work is supported in part by Grants-in-Aid of the Ministry
of Education, Culture, Sports, Science and Technology-Japan
 (Nos. 18104005, 
20340047,
20540248, 
21340049, 
22244018,  
and 24540250),
and Grant-in-Aid for Scientific Research on Innovative Areas
(No. 2004: 20105001, 20105002, 20105003, and 20105005).
\end{acknowledgments}

\bibliography{unsrt,apssamp}




\clearpage

\end{document}